\documentclass[twocolumn,preprintnumbers,amsmath,amssymb]{revtex4}

\usepackage{graphicx}
\usepackage{dcolumn}
\usepackage{bm}

\begin{document}

\preprint{YITP-04-18}

\title{$\theta$ vacuum effects on the chiral condensation 
and the $\eta^{\prime}$ meson correlators 
in the two-flavor massive QED$_2$ on the lattice}

\author{Hidenori Fukaya and Tetsuya Onogi\\}
\affiliation{
Yukawa Institute for Theoretical Physics,
             Kyoto University, Kyoto 606-8502, Japan
}
%


\begin{abstract}
We study the scalar and pseudoscalar condensations 
and the $\eta^{\prime}$ meson correlators of the two-flavor 
massive Schwinger model in the $\theta\neq 0$ vacuum. 
Exploiting our new method which was developed to investigate 
topological effects in the previous work, we find that 
$\langle\bar{\psi}\gamma_5\psi\rangle^{\theta} \neq 0$ and 
there exists a long-range correlation of the $\eta^{\prime}$ meson.
This phenomenon is well described by the clustering property. 
We also find that even in $\theta =0$ case the cancellation of the
long-range correlation is nontrivial and requires accurate 
contributions from higher topological sectors. 
Our results imply that the fluctuation of the ``disconnected'' diagram 
originates from the pseudoscalar condensation in each topological sector.  
\end{abstract}

\maketitle

\section{Introduction}
The topological structure is one of the essential aspects 
of the gauge theory. In addition to its  theoretical importance, 
it has phenomenological implications in particle physics 
such as the $\eta^{\prime}$ meson mass and the $\theta$ vacuum in QCD. 
Although much is known in the instanton dilute gas 
approximation, in order 
to go beyond the level of qualitative understanding
truly nonperturbative studies are required. 
For such purposes the lattice gauge theory should serve as 
a powerful tool, it has, however, been a nontrivial task to 
reproduce the topological structure since 
the naive lattice action does not preserve such properties.

Recently a new formulation of the lattice action, 
the Ginsparg-Wilson fermion \cite{Ginsparg:1981bj}, 
which realizes the exact chiral symmetry and the chiral anomaly has been proposed. 
Thus it would be an opportunity for the lattice gauge theory 
to provide a quantitative study of the topological structure 
of the gauge theories using this new formalism. 

In this work we study the $n_f=2$ massive Schwinger model
using this new lattice formalism 
and the method developed in our previous work~\cite{Fukaya:2003ph}.
From the analytical studies of this model 
in the continuum space \cite{Schwinger:tp, Coleman:1975pw, 
Coleman:1976uz, Smilga:1996pi, Hetrick:1995wq, Rodriguez:1996zj, 
Hosotani:1998za}, it has been found that the 
pion mass and the chiral condensates at strong coupling and small 
$\theta$ behave as
\begin{eqnarray}\label{eq:contconden}
m_{\pi}^{\theta} 
&=& c_{\pi} m^{2/3}g^{1/3}\cos^{2/3} \frac{\theta}{2}, \nonumber\\
- \langle \bar{\psi}\psi \rangle^{\theta} 
&=& c_{\bar{\psi}\psi} m^{1/3}g^{2/3}\cos^{4/3} \frac{\theta}{2}, \nonumber\\
i\langle \bar{\psi}\gamma_5\psi \rangle^{\theta} &=& 
c_{\bar{\psi}\psi} m^{1/3}g^{2/3}\sin  \frac{\theta}{2}\cos^{1/3} \frac{\theta}{2}, 
\end{eqnarray}
where $g$ denotes a gauge coupling constant,  $m$ is the fermion mass, 
$\gamma_5=$diag$(1,-1)$ and $c_{\pi}$, $c_{\bar{\psi}\psi}$ are numerical constants. 
Although there has been extensive lattice studies on the massive 
Schwinger model, they are limited to the scalar condensation 
$\langle \bar{\psi}\psi \rangle$ and the $\eta^{\prime}$ meson masses 
in the $\theta=0$ vacuum
\cite{Durr:2003xs,Durr:2000gi,Elser:2001pe,Elser:1996tb,
Kiskis:2000sb,Chandrasekharan:1998em,Gattringer:1995du,
deForcrand:1997fm,Vranas:1997da}.
 The lattice studies for the $\theta\neq 0$ 
vacuum done so far are only those in the pure U(1) gauge 
theory~\cite{Wiese:1988qz,Hassan:1994wy,Hassan:1995dn,
Imachi:1997hf,Plefka:1996tz,Azcoiti:2002vk,D'Elia:2003gr}.

Our goal is to compute the scalar and the 
pseudoscalar condensations $\langle \bar{\psi}\psi \rangle$,  
$\langle \bar{\psi}\gamma_5 \psi \rangle$ 
as well as the $\eta^{\prime}$ meson correlators 
$\langle \eta^{\prime\dagger} \eta^{\prime}  \rangle$ for fixed 
topological charges and study how the practical lattice 
calculations reproduce physical quantities in the $\theta$ vacuum.

In section \ref{sec:simulation},  we explain our method of simulations
and how to evaluate the $\theta$ vacuum effects by reweighting.
The results of the chiral condensation are presented in section
\ref{sec:cond}. The connection between the $\eta^{\prime}$ correlators and
the chiral condensations in each sector 
are shown in section \ref{sec:etaeach}.
In section \ref{sec:etamass},  we also present our result of 
the $\eta^{\prime}$ meson correlators in the $\theta$ vacuum.
The summary and the discussion are given in section \ref{sec:summary}.
Appendix is devoted to the study of the topological susceptibility. 

\section{Monte Carlo simulations and the reweighting}\label{sec:simulation}

In this section,  we briefly explain our method of the simulation. 
More details can be found in our previous paper \cite{Fukaya:2003ph}.
We take the gauge action proposed by L\"uscher \cite{Luscher:1998du} 
and the domain-wall fermion action~\cite{Kaplan,Shamir}, 
\begin{eqnarray}
S &=& \beta S_{G} + S_{F}, \\
S_{G}&=& \left\{\begin{array}{ll}
\displaystyle{\sum_{P}}\frac{(1-{\rm Re}P_{\mu\nu}(x))}
{1-(1-{\rm Re}P_{\mu\nu}(x))/\epsilon}
& \mbox{if admissible}
\\ \infty & \mbox{otherwise}
\end{array}
\right. , \label{eq:admaction2}\nonumber\\\\
S_{F} &=& \sum_{x, x^{\prime}}\sum_{s, s^{\prime}}
\sum^{2}_{i=1}
\left[
\bar{\psi}^{i}_{s}(x)D_{DW}(x, s ; x^{\prime}, s^{\prime})
\psi^{i}_{s^{\prime}}(x^{\prime})\right.
\nonumber\\
&&
\left. + \phi^{i\ast}_{s}(x)D_{AP}(x, s ; x^{\prime}, s^{\prime} )
\phi^{i}_{s^{\prime}}(x^{\prime})
\right], 
\end{eqnarray}
where $\beta=1/g^2$, $P_{\mu\nu}$ denotes the plaquette,  $\phi$'s are
Pauli-Villars regulators which cancel the bulk contribution of the
domain-wall fermions and $\epsilon$ is a fixed constant.
The gauge fields generated by this action satisfy L\"uscher's bound;
\begin{equation}
1-\mbox{Re}P_{\mu\nu}(x) < \epsilon\;\;\;\mbox{for all}\;\;x, \mu, \nu.
\end{equation}
Those gauge fields which satisfy the above condition are 
called ``admissible''. The ``admissible'' gauge fields have an exact 
topological charge defined as
\[
\displaystyle{Q \equiv \sum_x \frac{-i}{4\pi} 
\epsilon_{\mu\nu}\mbox{ln} P_{\mu\nu}},
\]
 on the lattice if $\epsilon < 2$. 
This charge is never changed 
in the evolution in the hybrid Monte Carlo algorithm so that 
we can evaluate the observables in each sector separately.
We take a $16\times16~ (\times 6)$ lattice at $\beta=1.0$ and
$\epsilon=1.0$. Fermion mass is chosen as $m=0.1, 0.15, 0.2, 0.25, 0.3$.
Fifty molecular dynamics steps with a step size $\Delta\tau=0.035$
are performed in one trajectory. Configurations are updated every 10 
trajectories. For each topological sector, around 500 trajectories 
are taken for the thermalization staring from the initial configuration 
which is the classical instanton solution with topological charge $Q$. 
We generate 300 configurations for each sector for the
measurements and from 1000 to 10000 for the reweighting factors at
various $\beta$. 

\begin{figure}[thb]
\includegraphics[width=8.5cm]{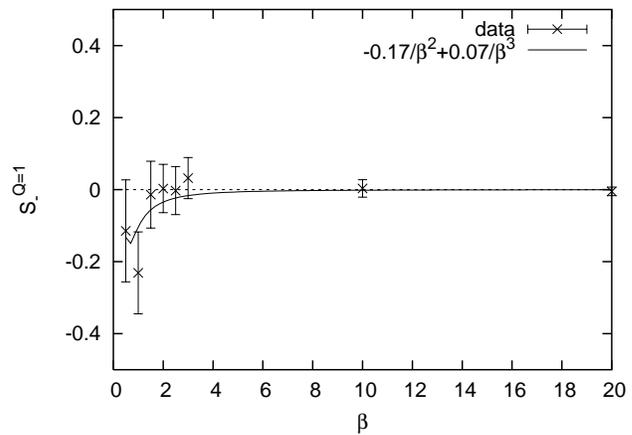}
\caption{
Plot of $\Delta S^{Q=1}(\beta, m=0.2)$. The solid line shows the result of
the fit. ($a_1=-0.17, a_2=0.07$, $\chi^2/\mbox{dof}=1.4$, 
$\int_{\beta=1.0}^{\infty} d\beta^{\prime}\Delta S^1(\beta^{\prime}, m)=-0.14(11)$).
}\label{fig:S-}
\end{figure}

\begin{figure}[htb]
\includegraphics[width=8.5cm]{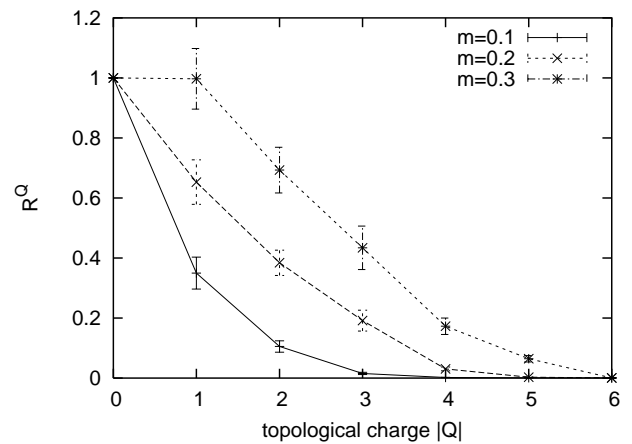}
\caption{
The reweighting factors at $m=0.1,0.2,0.3$ and $\beta=1.0$.
}
\label{fig:rewfac}
\end{figure}

\begin{figure}[htb]
\includegraphics[height=5cm]{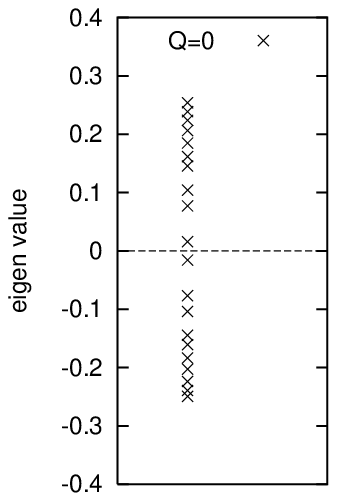}
\includegraphics[height=5cm]{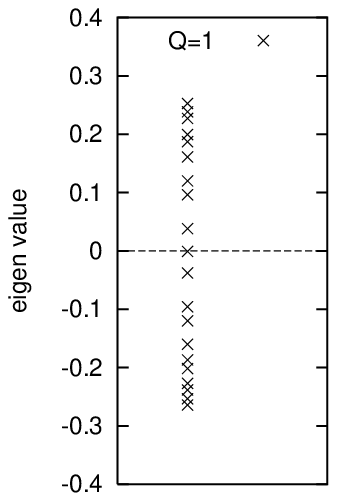}
\includegraphics[height=5cm]{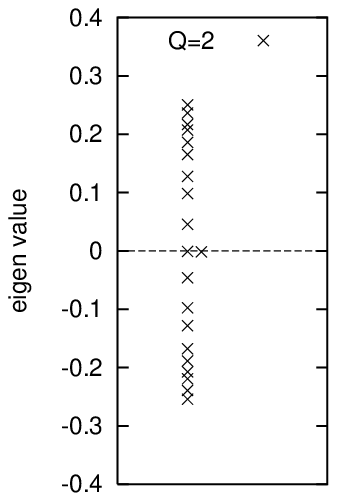}
\includegraphics[height=5cm]{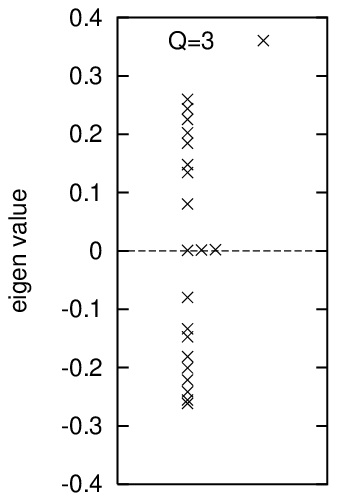}
\caption{
The eigenvalues of the domain-wall Dirac operator $\gamma_5D$
at $\beta=1.0$ in each sector. The index theorem is realized very well.
}
\label{fig:eigen}
\end{figure}

The expectation value of an operator $O$ in the $\theta$ vacuum 
is expressed as
\begin{equation}
\langle  O \rangle ^{\theta}_{\beta, m} = 
\frac{\sum_{Q=-\infty}^{+\infty}e^{iQ \theta}
       \langle  O \rangle^{Q}_{\beta, m}R^{Q}(\beta, m) }
     { \sum_{Q=-\infty}^{+\infty}e^{iQ \theta}R^{Q}(\beta, m) }, 
\end{equation}
where $\langle\rangle^Q_{\beta, m}$ denotes the expectation value
in the sector with $Q$ at $\beta, m$ and $R^Q$ is the reweighting 
factor;
\begin{eqnarray}
R^{Q}(\beta, m)&=&\frac{Z_{Q}(\beta, m)}{Z_{0}(\beta, m)}\nonumber\\
&=&e^{-\beta S^{Q}_{G\ \mbox{\tiny min}}}\times\mbox{Det}^Q
\times e^{\int_{\beta}^{\infty}d\beta^{\prime}\Delta S^Q(\beta^{\prime}, m)},
\end{eqnarray}
where $Z_Q$ is the partition function in each sector, 
$\mbox{Det}^Q$ denotes the contribution from fermion determinants
in the classical background with topological charge $Q$ and 
$S^Q_{G \mbox{\tiny min}}$ is the action of the background.
$\mbox{Det}^Q$ 's are calculated by the Householder method and the 
QL method. The integrals of $\Delta S^{Q}$ which is defined as
\begin{equation}\Delta S^Q(\beta^{\prime}, m)\equiv
\langle  S_{G} - S^{Q}_{G\ \mbox{\tiny min}}\rangle^{Q}_{\beta^{\prime}, m}
-\langle  S_{G} \rangle^{0}_{\beta^{\prime}, m},
\end{equation}
are evaluated by fitting the data with polynomials (See Fig.~\ref{fig:S-});
\begin{equation}
\Delta S^Q(\beta^{\prime}, m) = 
\frac{a_{1}}{\beta^{\prime \;2}}+\frac{a_{2}}{\beta^{\prime \;3}}.
\end{equation}
We present the $Q$ dependence of $R^Q$ in Fig.~\ref{fig:rewfac} 
and in Table~\ref{tab:rewfac}. 
The plots are consistent with the behavior indicated by 
the Atiyah-Singer index theorem. 
In Fig.\ref{fig:eigen} we show the lowest 20
eigenvalues with positive or negative chiralities 
of the massless Domain-wall Dirac operator for a typical 
configuration in each topological sector. 
The smallest eigenvalues are consistent with exact zero at $O(10^{-3})$, 
which suggests that the index theorem is realized very well 
due to the smallness of the violation of chiral symmetries.
As Fig.\ref{fig:rewfac} indicates,  we can ignore 
the contribution from large $Q$ sectors.
We evaluated the total expectation values by summing $-4\leq Q\leq 4$
sectors for $m=0.1, 0.15, 0.2$ and $-5\leq Q\leq 5$ sectors for $m=0.25, 0.3$.


\section{The chiral condensations}\label{sec:cond}
We measure the chiral condensations 
$\langle \bar{\psi}\psi\rangle^{Q}_{\beta, m}$ and 
$\langle \bar{\psi}\gamma_5\psi\rangle^{Q}_{\beta, m}$
with the fixed topological charge.  
The total expectation values of them in the $\theta$ vacuum 
can be obtained by the following formula,
\begin{eqnarray}
\langle \bar{\psi}\psi\rangle^{\theta}_{\beta, m} 
&=&
\frac{\sum_Q
e^{iQ\theta}\langle
\bar{\psi}\psi\rangle^{Q}_{\beta, m}R^Q(\beta, m)}
{\sum_Q
e^{iQ\theta}R^Q(\beta, m)},
\nonumber\\
i\langle \bar{\psi}\gamma_5\psi\rangle^{\theta}_{\beta, m} 
&=&
\frac{\sum_Q
e^{iQ\theta}i\langle
\bar{\psi}\gamma_5\psi\rangle^{Q}_{\beta, m}R^Q(\beta, m)}
{\sum_Q
e^{iQ\theta}R^Q(\beta, m)}.
\end{eqnarray}
In our practical calculation 
we truncate the sum over topological charge so that 
$Q$ is restricted to be $|Q| \leq Q_{\mbox{\tiny max}}$
where $Q_{\mbox{\tiny max}}=4$ for $m=0.1, 0.15, 0.2$ and 
$Q_{\mbox{\tiny max}}=5$ for $m=0.25, 0.3$. 
The scalar and the pseudoscalar operators are defined as
\begin{eqnarray}\label{eq:app}
\bar{\psi}\psi(x)&=&
\bar{\psi}(x, L_s)\psi(x, 0)+\bar{\psi}(x, 0)\psi(x, L_s), 
\nonumber\\
\bar{\psi}\gamma_5\psi(x)&=&
\bar{\psi}(x, L_s)\psi(x, 0)-\bar{\psi}(x, 0)\psi(x, L_s),
\end{eqnarray}
where $L_s=6$ denotes the size of the extra dimension 
of the domain-wall fermions.
In order to increase the statistics, spatial averages are 
always taken for the expectation values of local operators. 

\begin{figure*}[tph]
\includegraphics[width=8.5cm]{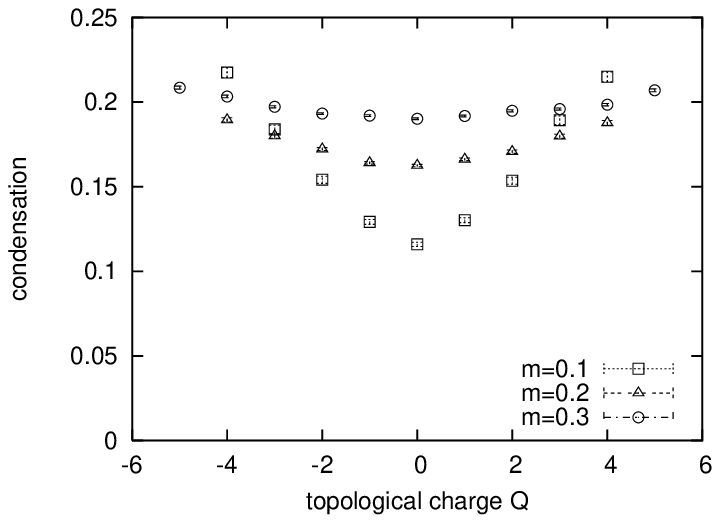}
\includegraphics[width=8.5cm]{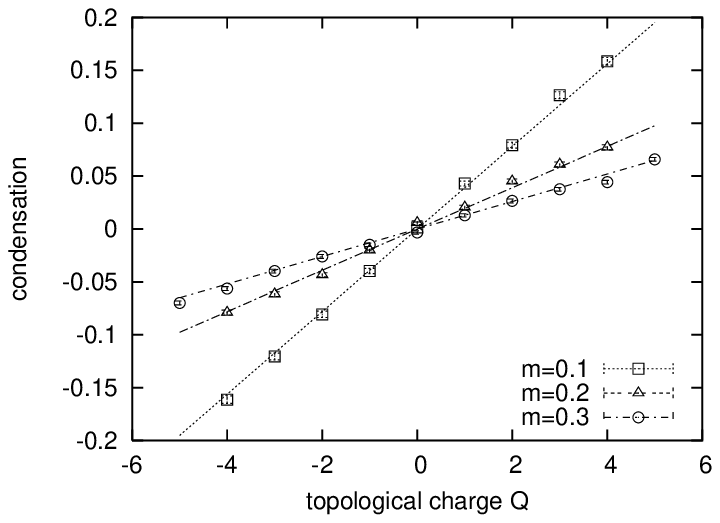}
\caption{
{\bf Left} :  Lattice results of the condensation 
$-\langle \bar{\psi}\psi \rangle^Q$ at $\beta=1.0$.
It increases with $|Q|$ symmetrically.
{\bf Right} : Lattice results of the condensation  
$-\langle\bar{\psi}\gamma_5\psi\rangle^Q$ 
in each topological sector at $\beta=1.0$.
The lines show the prediction from the anomaly equation
$-\langle\bar{\psi}\gamma_5\psi\rangle^Q = Q/(mV)$, which 
agrees with our data.}
\label{fig:condQ}
\includegraphics[width=8.5cm]{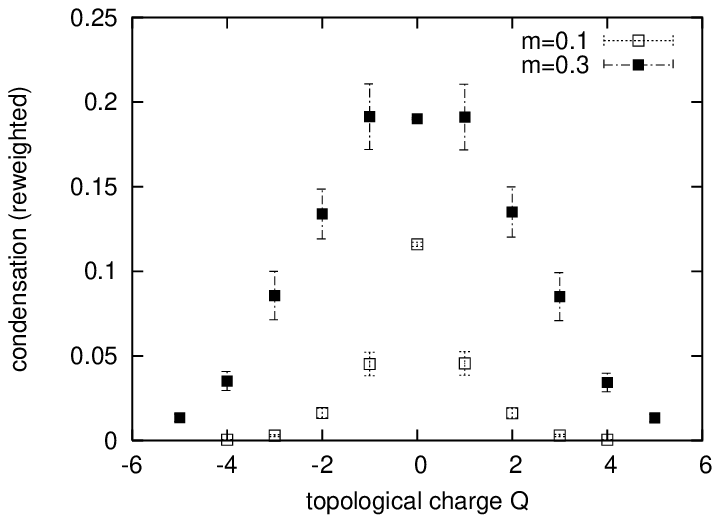}
\includegraphics[width=8.5cm]{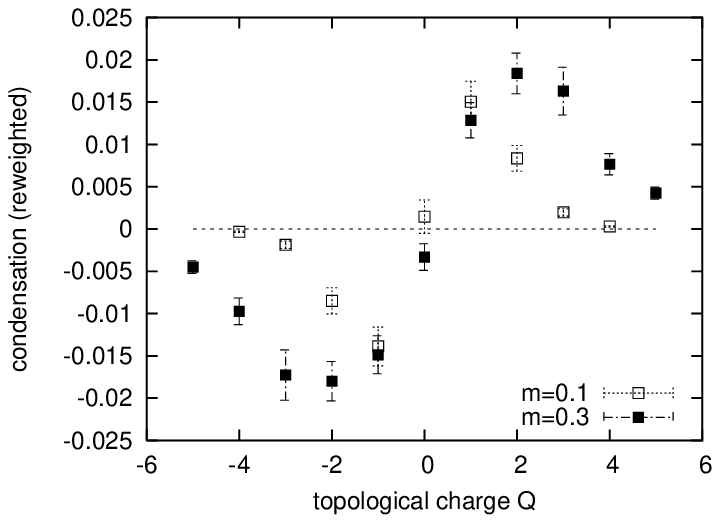}
\caption{
{\bf Left} : Reweighted condensation $-\langle \bar{\psi}\psi \rangle^QR^Q$ 
in each topological sector at $\beta=1.0$.
{\bf Right} : Reweighted condensation 
$-\langle\bar{\psi}\gamma_5\psi\rangle^QR^Q$ 
in each topological sector at $\beta=1.0$.
Note that the largest contribution comes from $Q\neq0$ sector.
}
\label{fig:condQrew}
\includegraphics[width=8.5cm]{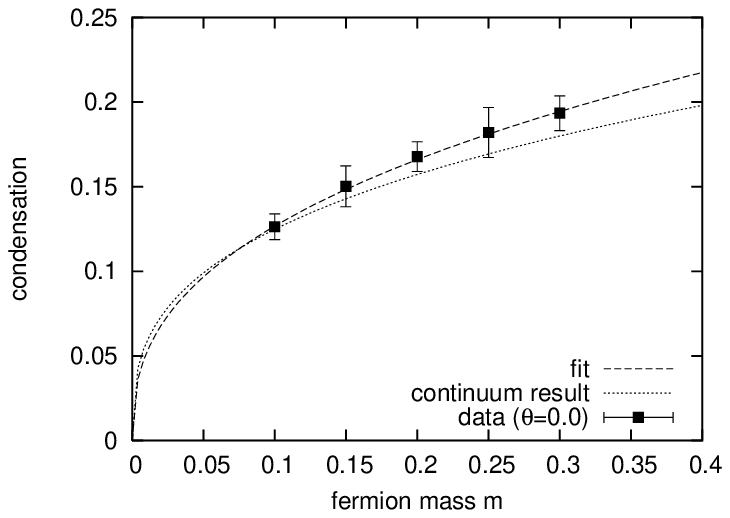}
\caption{
The fermion mass dependence of $-\langle \bar{\psi}\psi \rangle^{\theta}$
at $\beta=1.0$ and $\theta=0$.
The dashed line is the result of the fit with the function $Am^B$ 
($A=0.311(36),  B=0.388(68), \chi^2/\mbox{dof}=0.068$.)
The index $B$ is consistent with $B=1/3$ in the continuum theory.
The dotted line is the prediction of the condensation 
from the relation 
$-\langle \bar{\psi}\psi\rangle^{\theta} = c_{\bar{\psi}\psi}m_{\pi}^2/c_{\pi}^2m$
 with our data of $m_{\pi}$ in Table~\ref{tab:pimass} and 
$c_{\bar{\psi}\psi}/c_{\pi}^2=1/(4\pi)$ 
by Hetrick \textit{et al.}~ \cite{Hetrick:1995wq}.}
\label{fig:condfmass}
\end{figure*}

The condensations in each topological sector
$\langle \bar{\psi}\psi \rangle^Q$ and 
$\langle \bar{\psi}\gamma_5\psi \rangle^Q$ 
are plotted in Fig.\ref{fig:condQ}, where the numbers 
are given in Tables~\ref{tab:psipsi}, \ref{tab:psigammapsi}.
It is obvious that the $Q$ dependence of 
$\langle \bar{\psi}\psi\rangle^Q$ is symmetric and
that of $\langle \bar{\psi}\gamma_5\psi \rangle^Q$ is anti-symmetric, 
which can be understood from parity symmetry. 
The topological charge dependence of 
$\langle \bar{\psi}\gamma_5\psi \rangle^Q$ can be 
explained from the chiral anomaly equation given as 
\begin{eqnarray}
 \partial_{\mu} A^{\mu} 
&=& 2 m i \bar{\psi}\gamma_5\psi     + \frac{i}{2\pi}
\epsilon^{\mu\nu} F_{\mu\nu} .
\end{eqnarray}
Taking expectation values in the fixed topological sector 
and making summations over the whole spacetime volume,
one obtains the following relation
\begin{eqnarray}
- \langle \bar{\psi}\gamma_5\psi \rangle^Q &=&  Q/(m V),
\label{Anomaly}
\end{eqnarray}
which is in complete agreement with our lattice results.
We also plot the ``reweighted'' condensations
$\langle \bar{\psi}\psi \rangle^QR^Q$ and
$\langle \bar{\psi}\gamma_5\psi \rangle^QR^Q$
in Fig.\ref{fig:condQrew}.
It is important to note that most of the contributions 
in $i\langle\bar{\psi}\gamma_5\psi\rangle^{\theta\neq 0}$
come from $Q\neq 0$ sectors.  

The fermion mass dependence of the total condensations
$\langle \bar{\psi}\psi \rangle^{\theta=0} $
are presented in Fig.\ref{fig:condfmass} in 
which we also plot the fit function;
\begin{equation}
f(m)=Am^B.
\end{equation}
 Figure \ref{fig:condfmass} also shows the prediction from the analytic 
result of the continuum theory in Eq.(\ref{eq:contconden}) using 
$m_{\pi}^2$ as an input 
\begin{eqnarray}
 - \langle \bar{\psi}\psi \rangle &=& \frac{c_{\bar{\psi}\psi}}{c_{\pi}^2m} m_{\pi}^2.
\end{eqnarray}
In the following analysis we adopt the value of 
$c_{\bar{\psi}\psi}/c_{\pi}^2=1/(4\pi)$ by Hetrick \textit{et al.}\cite{Hetrick:1995wq}.
We find that our lattice results give a good agreement with the analytic 
results, as was also the case in the previous lattice 
studies ~\cite{Durr:2003xs,Durr:2000gi,Elser:2001pe,Elser:1996tb,
Kiskis:2000sb,Chandrasekharan:1998em,Gattringer:1995du,
deForcrand:1997fm,Vranas:1997da,Giusti:2001cn}.

\begin{figure*}[pth]
\includegraphics[width=8.5cm]{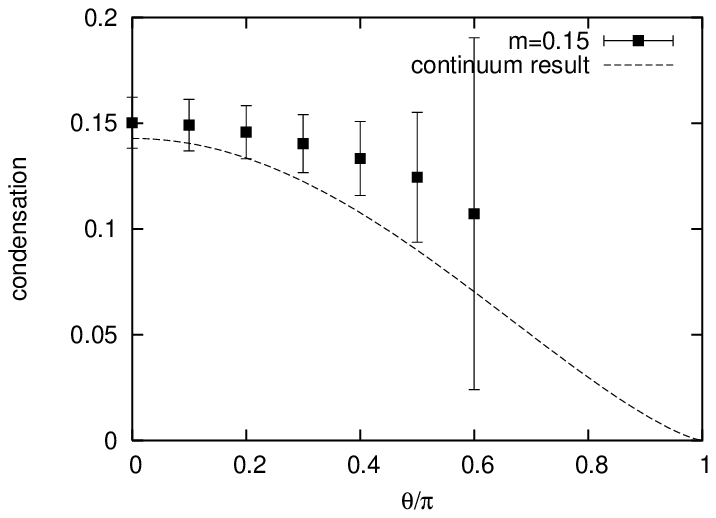}
\includegraphics[width=8.5cm]{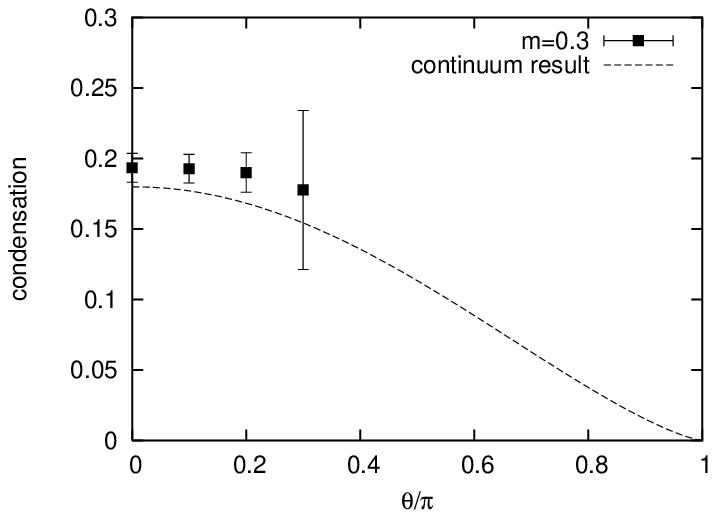}
\caption{
The $\theta$ dependence of $- \langle \bar{\psi}\psi \rangle^{\theta}$ 
at $\beta=1.0$ and $m=0.15, 0.3$. The dashed line shows the 
continuum result $A\cos^{4/3}\frac{\theta}{2}$ where $A=1/(4\pi) m_{\pi}^2$.
{\bf Left} : $m=0.15$. {\bf Right} : $m=0.3$.
}
\label{fig:condtheta}
\includegraphics[width=8.5cm]{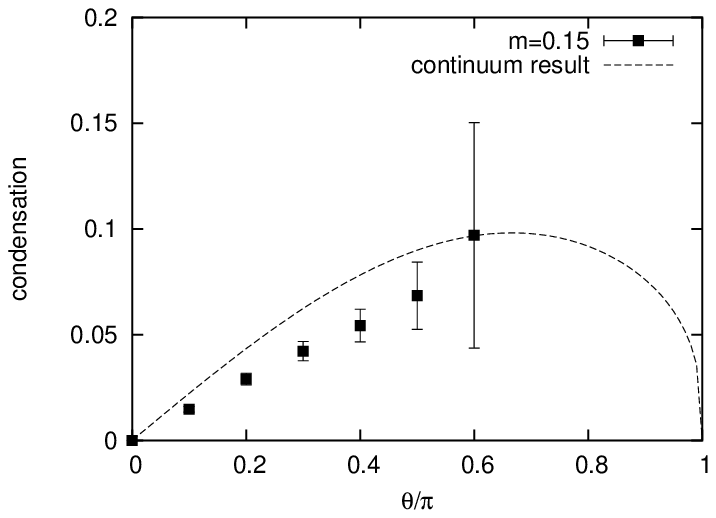}
\includegraphics[width=8.5cm]{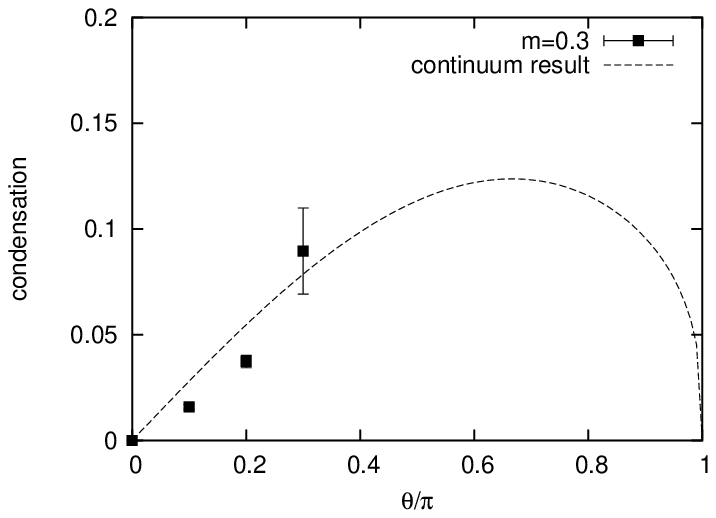}
\caption{
The $\theta$ dependence of $i\langle \bar{\psi}\gamma_5\psi \rangle^{\theta}$ 
at $\beta=1.0$ and $m=0.15, 0.3$. The dashed line shows the 
continuum result $A\sin \frac{\theta}{2}\cos^{1/3}\frac{\theta}{2}$
 where $A$ is the same value of $-\langle \bar{\psi}\psi \rangle^{\theta}$.
{\bf Left} : $m=0.15$. {\bf Right} : $m=0.3$.}
\label{fig:condaxtheta}
\end{figure*}

We also present the $\theta$ dependences of the total condensations at
 $m=0.15, 0.3$ as well as analytic results
in Fig.\ref{fig:condtheta} and Fig.\ref{fig:condaxtheta}.
The numerical data are given in Tables \ref{tab:psipsitheta}, 
\ref{tab:psigammapsitheta}. The qualitative features of the results 
in the continuum theory are realized in our simulation 
(See Eq.(\ref{eq:contconden})). Our lattice results show 30 \%
deviations from the analytic results. Possible source of this 
discrepancies may be the discretization error or the error 
in the reweighting factor.

%


\section{the $\eta^{\prime}$  meson correlators 
and the chiral condensations in each topological sector}\label{sec:etaeach}

The flavor singlet pseudoscalar meson $\eta^{\prime}$ 
is expected to be influenced by the topological properties 
of the vacuum. In QCD, the chiral anomaly gives a qualitative 
solution to the U(1) problem and it has been a long standing 
issue whether the mass gap between the flavor singlet 
and nonsinglet pseudoscalars can really be explained 
quantitatively from the lattice calculation. 
For this reason a number of unquenched lattice QCD simulations 
with $n_f=2$ to observe the mass 
gap~\cite{Venkataraman:1997xi,McNeile:2000hf,
Struckmann:2000bt,Lesk:2002gd} have been carried out.

The massive Schwinger model also has
the mass gap between the flavor singlet and 
nonsinglet pseudoscalars due to the chiral anomaly.
Since it is a two-dimensional theory,
the numerical cost is much smaller than QCD 
and the disconnected diagrams can be evaluated explicitly 
without relying on the noise method \cite{ref:noise}
or the Kuramashi's method 
\cite{ref:kuramashi}. 
The $\eta^{\prime}$ meson mass in $\theta=0$ vacuum 
has been obtained using the overlap fermion~\cite{Giusti:2001cn}.

The $\eta^{\prime}$ meson propagator is written as
\begin{eqnarray}
\langle \eta^{\prime\dagger}(x)\eta^{\prime}(y) \rangle 
\hspace{2in}\nonumber\\
=\left\langle(-\sum_{f=1}^2\bar{\psi}_f\gamma_5\psi_f(x))
(\sum_{f=1}^2\bar{\psi}_f\gamma_5\psi_f(y))\right\rangle
\hspace{.05in}\nonumber\\
=+2\left\langle \mbox{tr}\left( \gamma_5\frac{1}{D}(x, y)
\gamma_5\frac{1}{D}(y, x)\right)
\right\rangle \hspace{.4in}
\nonumber\\
-4 \left\langle 
\mbox{tr}\left(\gamma_5\frac{1}{D}(x, x)\right)
\mbox{tr}\left(\gamma_5\frac{1}{D}(y, y)\right)
\right\rangle,
\end{eqnarray}
where the  second term is so-called disconnected part.

\begin{figure*}[thbp]
\includegraphics[width=8.5cm]{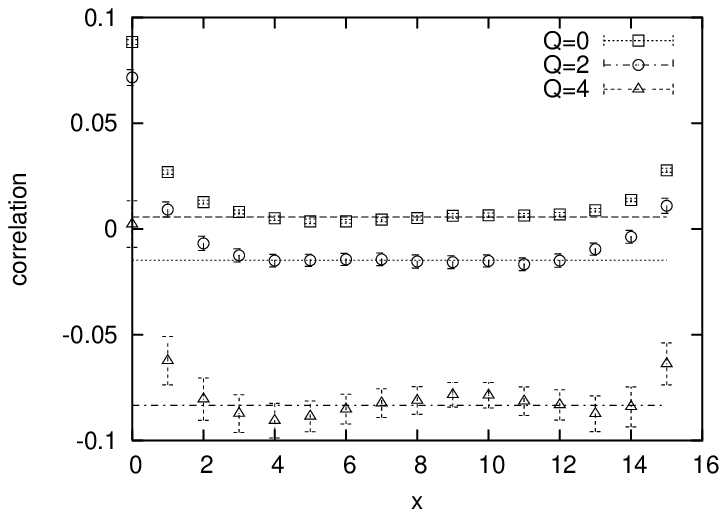}
\caption{
The propagators of the $\eta^{\prime}$ meson in each topological sector
at $\beta=1.0$ and $m=0.1$.
Long-range correlations are seen. The lines are the results of 
the fit with the function in Eq.(\ref{eq:etafit}).
}
\label{fig:etacorr}
\includegraphics[width=8.5cm]{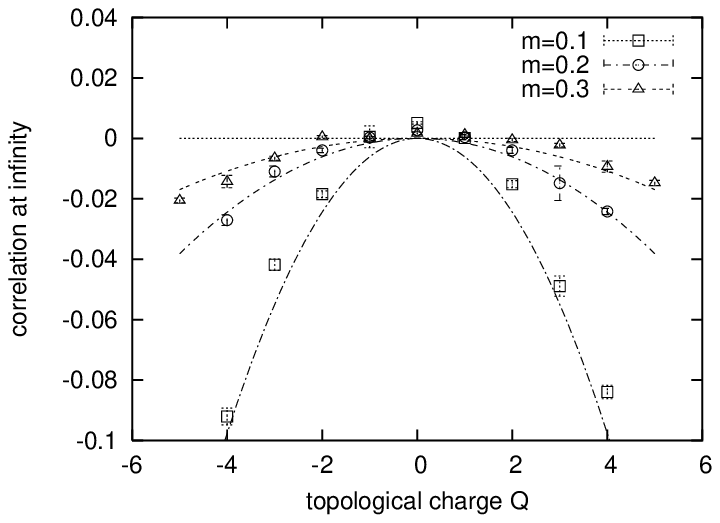}
\includegraphics[width=8.5cm]{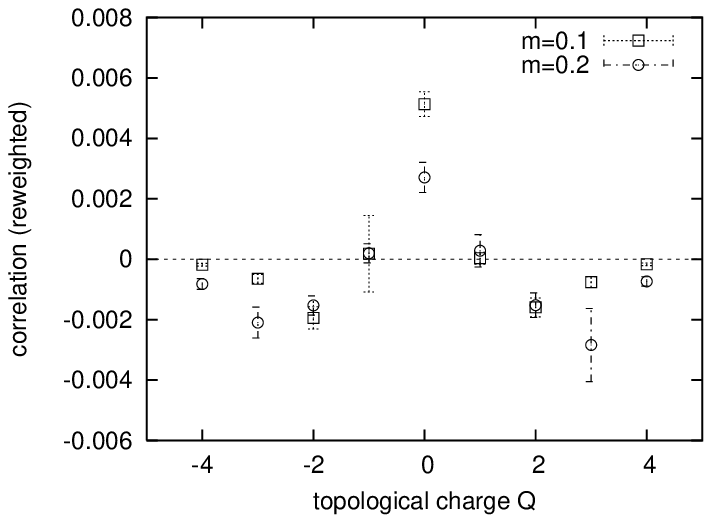}
\caption{{\bf Left} : 
The long-range correlations in each topological sector
$\langle \eta^{\prime\dagger}(\infty)\eta^{\prime}(0)\rangle^Q$.
The lines show the $Q$ dependence for large $Q$ derived 
from the clustering property 
$\langle \eta^{\prime\dagger}(\infty)\eta^{\prime}(0)\rangle^Q
= -4 Q^2/(mV)^2$, which agrees with our data quite well.  
{\bf Right} : The long-range correlations (reweighted)
$\langle \eta^{\prime}(\infty)\eta^{\prime}(0)\rangle^QR^Q$.
It is interesting that the long-range correlation of $Q=0$ sector
is non-zero and gives the largest contribution to the total expectation
value in the $\theta$ vacuum.
}
\label{fig:longQ}
\includegraphics[height=8.5cm, angle=-90]{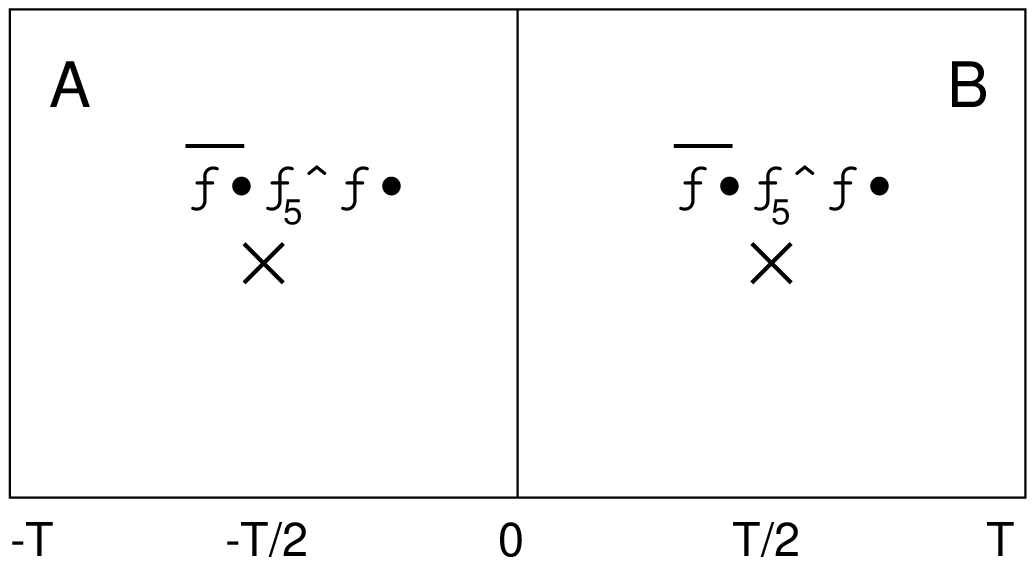}
\caption{
The long-range correlation of the $\eta^{\prime}$ meson 
can be understood as a summation of the products of the condensations 
which are determined by the topological charge 
in the region A and B respectively; 
$Z^Q_V\langle \eta^{\prime\dagger}(T/2)\eta^{\prime}(-T/2)\rangle^Q
\to_{T\to \infty}-\sum_{Q^{\prime}} 
Z^{Q^{\prime}}_{V_2} 
\langle \bar{\psi}_f\gamma_5\psi_f\rangle_B^{Q^{\prime}}
Z^{Q-Q^{\prime}}_{V_1} 
\langle \bar{\psi}_f\gamma_5\psi_f\rangle_A^{Q-Q^{\prime}}$}
\label{fig:etacorfig}
\end{figure*}

Fig.\ref{fig:etacorr} shows the propagators in each 
topological sector 
$\langle \eta^{\prime\dagger}(x)\eta^{\prime}(0)\rangle^Q$.
The result indicates the existence of long-range correlations.
We evaluate them by fitting the data with the function
\begin{equation}\label{eq:etafit}
f(x)=A(e^{-Bx}+e^{-B(L-x)})+C, 
\end{equation}
where $L=16$ is the size of the lattice.
We take $x\geq 3$ for the fitting range, we have also checked 
that the change of the value of $C$ for different fitting range
is small (less than 5\%). 

The data of $C$ in each topological sector are listed in Table \ref{tab:longcrr}. 
The results for $m=0.1,0.2,0.3$ are also plotted in Fig.\ref{fig:longQ}.
We find that there is a long-range correlation even in
$Q=0$ sector. Moreover, it is remarkable that it
gives the largest contribution to the total expectation value
in the $\theta$ vacuum.

We would like to explain this phenomenon by clustering properties.
Consider the two operators put on $t=T/2$ and $t=-T/2$ as 
Fig.\ref{fig:etacorfig}.  
We expect that in the large volume limit the path integrals 
for the correlation functions in the volume $V$ can be 
expressed in terms of the chiral condensations as follows, 
\begin{eqnarray}
Z^{Q}_V
\left\langle \eta^{\prime\dagger}(T/2)\eta^{\prime}(-T/2)\right\rangle^Q
\stackrel{T\to\infty}{\to}
\hspace{1.2in}
\nonumber\\\nonumber\\
-\sum_{Q^{\prime}} 
Z^{Q^{\prime}}_{V_2}
\langle \sum_f\bar{\psi}_f\gamma_5\psi_f
\rangle^{Q^{\prime}}_B
Z^{Q-Q^{\prime}}_{V_1}
\langle \sum_f\bar{\psi}_f\gamma_5\psi_f
\rangle^{Q-Q^{\prime}}_A,
\end{eqnarray}
where $Z$'s denote the partition functions 
and $\langle\rangle^{Q^{\prime}}_{A, B}$ means
the expectation value in the region $A, B$ with volume $V_1,V_2$
($V=V_1+V_2$) in which the gauge fields have a topological 
charge $Q^{\prime}$ in that region.
In $Q=0$ case taking $V_1=V_2=V/2$ we obtain
\begin{eqnarray}
Z^{0}_V
\left\langle \eta^{\prime\dagger}(T/2)\eta^{\prime}(-T/2)\right\rangle^{Q=0}
\stackrel{T\to\infty}{\to}
\hspace{1in}
\nonumber\\\nonumber\\
-\sum_{Q^{\prime}} Z^{Q^{\prime}}_{V/2}
\langle \sum_f\bar{\psi}_f\gamma_5\psi_f\rangle^{Q^{\prime}}_B
Z^{-Q^{\prime}}_{V/2}
\langle \sum_f\bar{\psi}_f\gamma_5\psi_f\rangle^{-Q^{\prime}}_A
\nonumber\\\nonumber\\
=\;+\sum_{Q^{\prime}} (Z^{Q^{\prime}}_{V/2})^2
\left(\langle \sum_f\bar{\psi}_f\gamma_5\psi_f
\rangle^{Q^{\prime}}_A\right)^2>0, \hspace{.55in}
\end{eqnarray}
where we assume $Z^Q_{V/2}=Z^{-Q}_{V/2}$, 
$\langle O\rangle_A^Q=\langle O\rangle_B^Q$
and we use the anti-symmetry; 
\begin{equation}
\langle \bar{\psi}\gamma_5\psi\rangle^Q
=-\langle \bar{\psi}\gamma_5\psi\rangle^{-Q}, 
\end{equation}
as seen in Eq.(\ref{Anomaly}) and Fig.\ref{fig:condQ}.
Moreover, the $Q$ dependence of the long-range correlations 
for large $Q$ and the large volume limit can also be understood. 
In this limit, since $Q$, $V$ are extensive quantities, 
the free energy $F(Q,V)$ defined as $F=-\mbox{ln}(Z^{Q}_V)$ which is 
another extensive quantity should be expressed as 
\begin{eqnarray}
F &=& V  f\left(\frac{Q}{V}\right),
\label{free_enegy}
\end{eqnarray}
where $f$ is some unknown function.
Substituting Eq.(\ref{free_enegy}) and approximating 
the sum over $Q^{\prime}$ by the integral over $Q^{\prime}$,
we obtain 
\begin{eqnarray}
Z^{Q}_V
\left\langle \eta^{\prime\dagger}(T/2)\eta^{\prime}(-T/2)\right\rangle^Q
\stackrel{T\to\infty}{\to}
\hspace{1in}
\nonumber\\\nonumber\\
-\int dQ^{\prime} 
\frac{2(Q-Q^{\prime})}{m V_1}
\frac{2Q^{\prime}}{m V_2}
e^{ -( V_1 f(\frac{Q-Q^{\prime}}{V_1})
       + V_2 f(\frac{Q^{\prime}}{V_2} ) )},
\label{longcorr_calc}
\end{eqnarray}
where we use Eq.(\ref{Anomaly})~(see also Fig.\ref{fig:condQ}).
In the large volume limit, the integral over $Q^{\prime}$ 
is dominated by $Q_{\ast}^{\prime}=\frac{Q V_2}{V_1+V_2}$, 
which minimizes the total free energy. We can evaluate the 
integral by integrating the fluctuations of $Q^{\prime}$ 
around $Q_{\ast}^{\prime}$. Changing variables as 
$Q^{\prime}=Q_{\ast}^{\prime}+q$ and expanding the free energy in $q$
to the second order,
right hand side of Eq.(\ref{longcorr_calc}) becomes
\begin{equation}
-\int dq
\frac{4}{m^2} \left(\frac{Q^2}{V^2}-\frac{q^2}{V_1 V_2}\right)
e^{ -( V f(\frac{Q}{V})
+\frac{V q^2}{2V_1 V_2} f^{\prime\prime}(\frac{Q}{V}) )},
\end{equation}
where $V=V_1+V_2$.
Performing the integral over $q$ to the first order we obtain
\begin{equation}
\left\langle \eta^{\prime\dagger}(T/2)\eta^{\prime}(-T/2)\right\rangle^Q
 \stackrel{T\to\infty}{\to}
- \frac{4}{m^2}\left(\frac{Q^2}{V^2}- 
\frac{1}{V f^{\prime\prime}(\frac{Q}{V})}\right).
\label{lrcorr_prediction}
\end{equation}
The second term is indeed suppressed in the limit 
where both $Q$ and the volume becomes large as expected.
In Fig.\ref{fig:longQ}, we compare our lattice results 
on the $Q$ dependence of the long-range correlation and the first 
term in Eq.(\ref{lrcorr_prediction}).  It is surprising that this 
argument describes the results in Fig.\ref{fig:longQ} very well. 

Thus we conclude that there is a relation between 
the long-range correlation of $\eta^{\prime}$  and the chiral
condensations, which is understood  by the clustering property 
of the theory based on the discussion of $Q^{\prime}$ instantons in
the half of the space and $Q^{\prime}-Q$ anti-instantons in the another
half. Here we would like to emphasize two points;
\begin{itemize}
 \item The $\eta^{\prime}$ meson has a long-range correlation 
       in each topological sector
       even in $Q=0$ case in spite of the fact that
       $\langle\bar{\psi}\gamma_5\psi\rangle^{Q=0}$ vanishes. 
 \item Our lattice data at large $Q$ satisfy the following relation,
       \begin{eqnarray}
	 \langle \eta^{\prime\dagger}(\infty )\eta^{\prime}(0)\rangle^Q_V 
        & \sim & - \frac{4}{m^2}\frac{Q^2}{V^2},
	\end{eqnarray}
        which can be explained by the clustering property.
\end{itemize}

\section{The $\eta^{\prime}$ meson in the $\theta$ vacuum}\label{sec:etamass}

\begin{figure*}[pth]
\includegraphics[width=8.5cm]{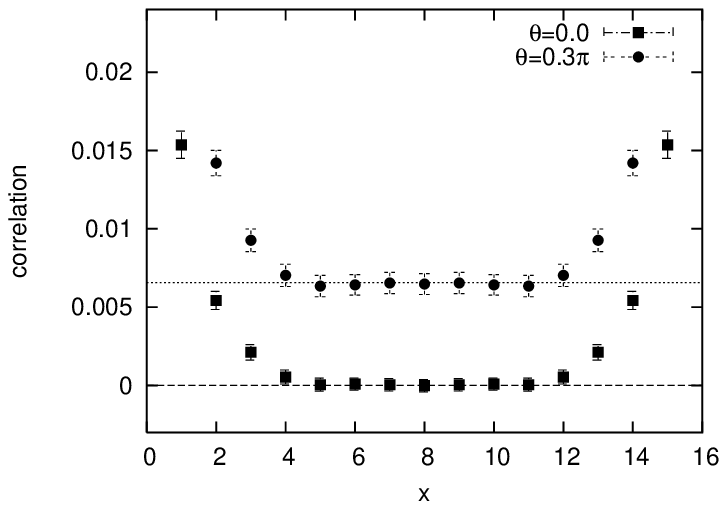}
\includegraphics[width=8.5cm]{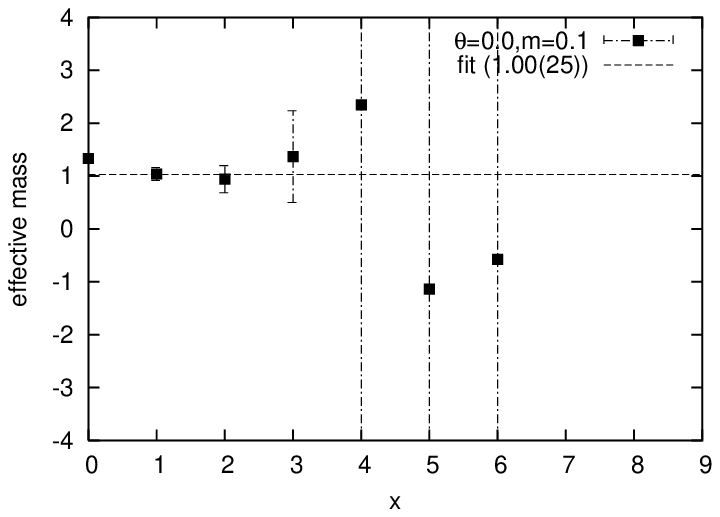}
\caption{{\bf Left} : 
The propagations of the $\eta^{\prime}$ meson at $\theta=0, 0.3\pi$ and
 $m=0.1$ are shown. The lines show $C_{\theta}$
in Eq.(\ref{eq:etatheta}). It is obvious that 
there are long-range correlations at $\theta\neq 0$.
On the other hand,  $\theta=0$ case is consistent with zero.\\
{\bf Right} : 
The effective mass plot of the $\eta^{\prime}$ meson at $\theta=0$ and $m=0.1$.
The dashed line shows the result of the fit in Eq.(\ref{eq:etatheta}).
The fitting range is $x\geq 2$.
}
\label{fig:etaproptheta}
\includegraphics[width=8.5cm]{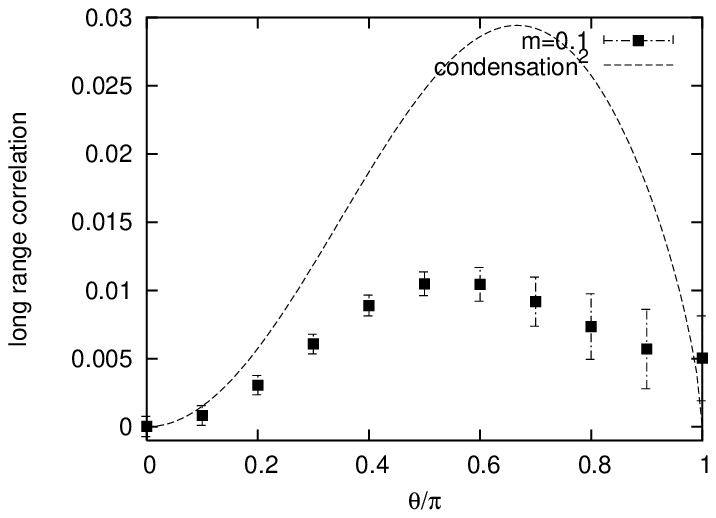}
\caption{
The long-range correlations of the $\eta^{\prime}$ meson
in the $\theta$ vacuum 
$\langle \eta^{\prime\dagger}(\infty )\eta^{\prime}(0)\rangle^{\theta}$.
The dashed line shows the result of 
$4|\langle\bar{\psi}\gamma_5\psi\rangle^{\theta}|^2$ in the continuum theory.
}
\label{fig:longcrrtheta}
\includegraphics[width=8.5cm]{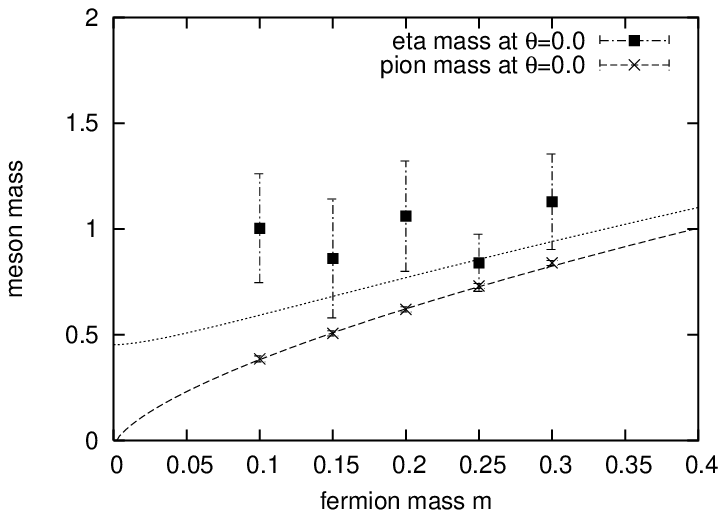}
\includegraphics[width=8.5cm]{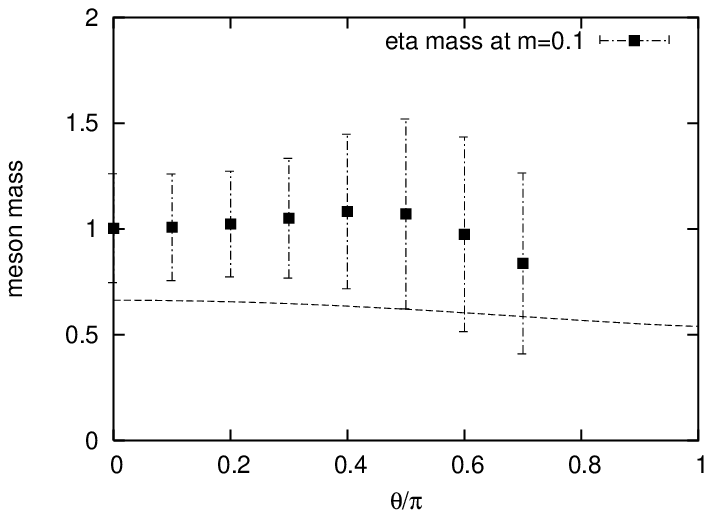}
\caption{{\bf Left} : 
The $\eta^{\prime}$ meson mass and the pion mass at $\theta=0$.
The lines show the continuum results from Hosotani et.al 
\cite{Hosotani:1998za} in which the effective coupling constant
$g_{\mbox{eff}}=0.68$ is determined from the pion mass at $m=0.1$ and $\theta =0$.
{\bf Right} : 
The $\theta$ dependence of the $\eta^{\prime}$ meson at $m=0.1$.
}
\label{fig:etamass}
\end{figure*}

From the discussion in section \ref{sec:etaeach}, we expect 
that the correlation
of the $\eta^{\prime}$ meson in the $\theta$ vacuum is expressed 
by the following function, 
\begin{equation}\label{eq:etatheta}
\langle \eta^{\prime\dagger}(x)\eta^{\prime}(0)\rangle^{\theta}
=A(e^{-m_{\eta^{\prime}}x}+e^{-m_{\eta^{\prime}}(L-x)})+C_{\theta}.
\end{equation}
We plot the data of 
$\langle\eta^{\prime\dagger}(x)\eta^{\prime}(0)\rangle^{\theta}$ 
at $\theta=0, 0.3 \pi,m=0.1$
in the left of Fig.\ref{fig:etaproptheta}.
Fitting the data with Eq.(\ref{eq:etatheta}), we evaluate
the $\eta^{\prime}$ meson mass and the long-range correlation $C_{\theta}$.
The fitting range is determined as $x>2$ from the effective mass 
at $\theta=0$ (See Fig.\ref{fig:etaproptheta}).

As Fig.\ref{fig:longcrrtheta} and Table \ref{tab:longcrrtheta}
show, the qualitative feature of
the long-range correlations is consistent with 
$4|\langle\bar{\psi}\gamma_5\psi\rangle^{\theta}|^2$ at
small $m$. Especially it is remarkable that our data
show the cancellation of the long-range correlations in the
$\theta =0$ vacuum at $m=0.1,0.15$.
On the other hand, we find that it is difficult to evaluate 
them precisely as $m$ increases since the $Q$ dependence the operator
is steep ($\propto Q^2$ ).  In fact, 
the long-range correlation at $m=0.2$ gives 3.5 $\sigma$ 
deviation from zero. This is because one gets contributions 
from larger number of topological sectors as shown in 
Fig.\ref{fig:longQ} so that the cancellation becomes more delicate. 
Obviously the present level of accuracies for 
the reweighting factors $R^Q$ and the condensation 
$\langle \bar{\psi} \gamma_5 \psi \rangle$ are not sufficient 
for larger masses.  Moreover, at $m = 0.25 , 0.3$ the sectors 
with $|Q| > |Q_{\rm max}|$ cannot be ignored. 
In order to have a good control of the long-range correlation
for larger masses, one has to reduce the statistical or 
systematic errors of 
$R^Q$ and $\langle \bar{\psi} \gamma_5 \psi \rangle$,
while also evaluating higher topological sectors.

The data of the $\eta^{\prime}$ meson mass are presented in Fig.\ref{fig:etamass}.
They have large statistical errors but 
we find that the $\eta^{\prime}$ meson is indeed heavier than the pion
as predicted by the continuum theory.

\section{Summary and discussion}\label{sec:summary}

We investigate the chiral condensations and the $\eta^{\prime}$ 
meson propagators in each topological sector and in the $\theta$ vacuum.
Let us summarize the results as follows.
\begin{enumerate}
 \item The chiral condensations are nonzero in each sector, 
       \begin{eqnarray}
        -\langle\bar{\psi}\psi\rangle^Q\neq 0 \;\;(\mbox{Q symmetric})\nonumber\\
	-\langle\bar{\psi}\gamma_5\psi\rangle^Q=\frac{Q}{m V} 
        \neq 0 \;\;(\mbox{Q anti-symmetric}), 
       \end{eqnarray}
       where the second equation is consistent with the anomaly equation.
       The total expectation values in the $\theta$ vacuum obtained
       by the reweighting method are qualitatively consistent with 
       the results of the  continuum theory.
 \item There actually exist the long-range correlations of 
       the $\eta^{\prime}$ meson  in each sector, 
       \begin{equation}
	\langle\eta^{\prime\dagger}(\infty)\eta^{\prime}(0)\rangle^Q\neq 0, 
       \end{equation}
       even in $Q=0$ case. It is also remarkable that 
       our data show the following $Q$ dependence of the long-range 
       correlation as 
       \begin{eqnarray}
	\langle\eta^{\prime\dagger}(\infty)
	 \eta^{\prime}(0)\rangle^{Q=0}>0, \nonumber\\
	\langle\eta^{\prime\dagger}(\infty)
	 \eta^{\prime}(0)\rangle^{Q>2}\sim -\frac{4Q^2}{m^2V^2}<0, 
       \end{eqnarray}
       which can be explained by the clustering property. 
       This means that the precise measurements of higher topological
       sectors are necessary to cancel the long-range correlation 
       in the total expectation value at $\theta=0$.
 \item Thus the $\eta^{\prime}$ meson correlations in the $\theta$
       vacuum should be treated as the following function, 
       \begin{eqnarray}
	\langle\eta^{\prime\dagger}(x)\eta^{\prime}(0)
	 \rangle^{\theta}=A(e^{-m_{\eta^{\prime}}x}
	 +e^{-m_{\eta^{\prime}}(L-x)})
	 +C_{\theta},\hspace{-0.5in}\nonumber\\ 
       \end{eqnarray}
       even in the $\theta=0$ vacuum.
       Our results of $C_{\theta}$ and $m_{\eta^{\prime}}$ from 
       the fit are qualitatively consistent with the continuum
       results as $\langle\eta^{\prime\dagger}(\infty)
        \eta^{\prime}(0)\rangle^\theta\propto
        |\langle\bar{\psi}\gamma_5\psi\rangle^\theta|^2$, 
       although there are discrepancies at the quantitative level.
\end{enumerate}
It is now clear why the contribution from ``disconnected'' diagrams
is very noisy. It is because each configuration gives 
the pseudoscalar condensation. 
They must be canceled at $\theta=0$ by parity symmetry.
We should however take the existence into account since our
simulation would have both of the systematic and statistical errors
in the evaluation of the higher topological sectors.
Note that 4-dimensional QCD might have the same problems.
It would be also important to investigate the topological effects on
the $\bar{\psi}\gamma_5\psi$ condensation and the $\eta^{\prime}$ meson
propagations in QCD.

It should be stressed that the present lattice formalism 
of the chiral fermion  already allows us to study 
the topological effects as a well defined problem in principle. 
Although our method may be limited to two dimensions, 
it would be important to investigate some new efficient method 
or algorithm for the study of the topological effect in QCD.

\section*{ACKNOWLEDGMENTS}\label{sec:acknowlegments}
The authors would like to thank H.~Matsufuru, T.~Umeda, 
Y.~Kikukawa, S.Hashimoto, Y.Aoki , K.~Takahashi and 
T.~Takimi for useful discussions and comments. 
The author also express special thanks to 
T.~Izubuchi for crucial discussions on the relation 
between the chiral condensation and the anomaly equation.
They would also like to acknowledge  M. Hamanaka and 
S.~Sugimoto for informing us of the analytic results 
on instantons and the massive Schwinger model.
The authors thank the Yukawa Institute for Theoretical Physics 
at Kyoto University, where this work was initiated during the 
YITP-W-02-15 on ``YITP School on Lattice Field Theory''.
The numerical simulations were done on NEC SX-5 at Research  
Center for Nuclear Physics in Osaka University, and Hitachi 
SR8000 model F1 supercomputer at KEK.
This work was supported in part by Grant-in-Aid for Scientific Research 
from the Ministry of Education, Culture, Sports, Science and Technology of 
Japan (No.13135213).

They would also like to thank to YITP members for constant 
encouragements.

\appendix
\section*{Appendix}\label{sec:Appendix}
In this appendix, we present our results of the topological
susceptibility $\chi$ defined as
\begin{eqnarray}
\chi \equiv \frac{1}{V}\langle Q^2 \rangle .
\end{eqnarray}
The topological susceptibility is a useful measure of the topological 
fluctuations of the vacuum, which is extensively studied 
in QCD \cite{Gausterer:1988tg,Bitar:1991wr,Kuramashi:1993mv,Hart:2001pj,
 Bali:2001gk,Hasenfratz:2001wd,Alles:2000cg,Bernard:2003gq,:2004ij}.

\begin{figure}[h]
\includegraphics[width=8.5cm]{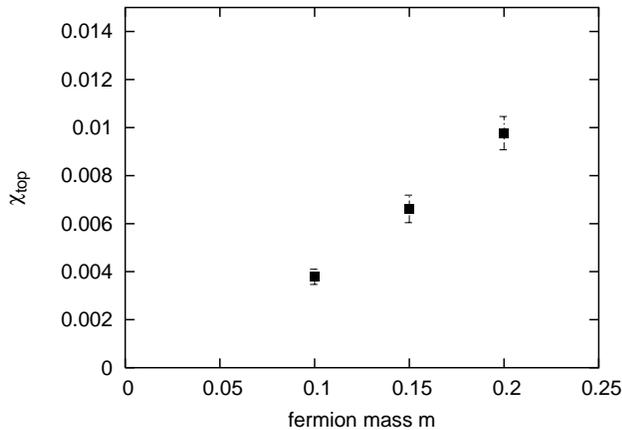}
\caption{
The mass dependence of the topological susceptibility $\chi$
} 
\label{fig:chi}
\end{figure}
Fig.\ref{fig:chi} indeed shows a mass dependence of $\chi$ which 
decreases towards the chiral limit somewhat faster than linearly 
in $m$. It would be interesting to 
compare it  with the analytical results. Recently, the topological 
susceptibility was also studied using the overlap fermion and 
staggered fermion \cite{Durr:2003xs}, where they find $\chi$ 
decreases slower than linearly in $m$ when the  Leutwyler-Smilga 
parameter $x\equiv V\Sigma m$ is large.

\newpage


\begin{table*}[p]
\begin{tabular}{cccccc}
\hline
\hline
 & $m=0.1$ & $m=0.15$ & $m=0.2$ & $m=0.25$ & $m=0.3$ \\\hline
 $R^0$  &   1.000(0)  & 1.000(0)  & 1.000(0)    & 1.000(0)  & 1.000(0)\\
 $R^1$  &   0.349(53) & 0.74(12)  & 0.652(74)   & 0.78(14)  & 0.99(10) \\ 
 $R^2$  &   0.105(19) & 0.295(47) & 0.384(42)   & 0.63(11)  & 0.693(76)\\
 $R^3$  &   0.0154(34)& 0.060(10) & 0.191(34)   & 0.285(51) & 0.434(72)\\
 $R^4$  &   0.00199(51)&0.0166(46)& 0.0304(62)  & 0.129(25) & 0.173(27)\\
 $R^5$  &             &           & 0.00317(63) & 0.0346(70)&
 0.064(11)\\
\hline\hline
\end{tabular}
\caption{
The reweighting factors $R^Q(\beta,m)$ at $\beta=1.0$ for various topological 
sectors and the fermion masses.}
\label{tab:rewfac}
\end{table*}

\begin{table*}[p]
\begin{tabular}{lc}
\hline
\hline
 & pion mass \\\hline
$m=0.1$     & 0.386(14)\\
$m=0.15$    & 0.507(13)\\
$m=0.2$     & 0.620(12)\\
$m=0.25$    & 0.730(12)\\
$m=0.3$     & 0.840(12)\\
\hline\hline
\end{tabular}
\caption{The pion mass at $\beta=1.0$ and $\theta=0$.}
\label{tab:pimass}
\end{table*}

\begin{table*}[p]
\begin{tabular}{lccccc}
\hline
\hline
 & $m=0.1$ & $m=0.15$ & $m=0.2$ & $m=0.25$ & $m=0.3$ \\\hline
 $-\langle\bar{\psi}\psi\rangle^0$
   &   0.1160(13) & 0.14324(85) & 0.16251(68) & 0.17861(56) & 0.19018(48)\\
 $-\langle\bar{\psi}\psi\rangle^1$   
   &   0.1302(14) & 0.14709(99) & 0.16619(78) & 0.17839(60) & 0.19181(52)\\
 $-\langle\bar{\psi}\psi\rangle^{-1}$
   &   0.1292(14) & 0.14852(95) & 0.16407(71) & 0.17992(65) & 0.19200(51)\\
 $-\langle\bar{\psi}\psi\rangle^2$
   &   0.1535(20) & 0.1591(12)  & 0.17068(97) & 0.18208(68) & 0.19490(58)\\
 $-\langle\bar{\psi}\psi\rangle^{-2}$
   &   0.1541(21) & 0.1611(14)  & 0.17223(84) & 0.18314(63) & 0.19326(55)\\
 $-\langle\bar{\psi}\psi\rangle^3$
   &   0.1894(26) & 0.1758(15)  & 0.1798(12)  & 0.18760(87) & 0.19592(62)\\
 $-\langle\bar{\psi}\psi\rangle^{-3}$
   &   0.1839(25) & 0.1805(18)  & 0.1800(13)  & 0.19051(89) & 0.19725(62)\\
 $-\langle\bar{\psi}\psi\rangle^4$
   &   0.2150(30) & 0.1912(18)  & 0.1877(15)  & 0.19526(98) & 0.19848(68)\\
 $-\langle\bar{\psi}\psi\rangle^{-4}$
   &   0.2175(29) & 0.1933(18)  & 0.1895(13)  & 0.19429(98) & 0.20334(78)\\
 $-\langle\bar{\psi}\psi\rangle^5$
   &              &             &             & 0.2040(12)  & 0.20702(83)\\
 $-\langle\bar{\psi}\psi\rangle^{-5}$
   &              &             &             & 0.2030(11)  & 0.20857(87)\\
\hline\hline
\end{tabular}
\caption{ The scalar condensation 
$-\langle\bar{\psi}\psi\rangle^Q$ for a fixed topological charge $Q$ 
at $\beta=1.0$.}
\label{tab:psipsi}
\end{table*}

\begin{table*}[p]
\begin{tabular}{lccccc}
\hline
\hline
 & $m=0.1$ & $m=0.15$ & $m=0.2$ & $m=0.25$ & $m=0.3$ \\\hline
 $-\langle\bar{\psi}\gamma_5\psi\rangle^0$
   &   0.0014(20) & 0.0001(19) & 0.0058(18) & 0.0045(19) &-0.0033(16)\\
 $-\langle\bar{\psi}\gamma_5\psi\rangle^1$   
   &   0.0430(23) & 0.0263(22) & 0.0205(20) & 0.0125(18) & 0.0129(16)\\
 $-\langle\bar{\psi}\gamma_5\psi\rangle^{-1}$
   &  -0.0396(25) &-0.0288(20) &-0.0200(19) &-0.0181(18) &-0.0149(17)\\
 $-\langle\bar{\psi}\gamma_5\psi\rangle^2$
   &   0.0794(28) & 0.0528(22) & 0.0452(20) & 0.0303(18) & 0.0265(17)\\
 $-\langle\bar{\psi}\gamma_5\psi\rangle^{-2}$
   &  -0.0807(30) &-0.0584(23) &-0.0433(20) &-0.0349(17) &-0.0260(18)\\
 $-\langle\bar{\psi}\gamma_5\psi\rangle^3$
   &   0.1266(33) & 0.0834(25) & 0.0611(23) & 0.0469(20) & 0.0376(17)\\
 $-\langle\bar{\psi}\gamma_5\psi\rangle^{-3}$ 
   &  -0.1205(32) &-0.0896(26) &-0.0616(24) &-0.0524(20) &-0.0398(17)\\
 $-\langle\bar{\psi}\gamma_5\psi\rangle^4$
   &   0.1586(36) & 0.1067(27) & 0.0773(23) & 0.0670(19) & 0.0443(18)\\
 $-\langle\bar{\psi}\gamma_5\psi\rangle^{-4}$
   &  -0.1613(36) &-0.1094(28) &-0.0789(23) &-0.0637(20) &-0.0563(18)\\
 $-\langle\bar{\psi}\gamma_5\psi\rangle^5$
   &              &            &            & 0.0828(22) & 0.0659(17)\\
 $-\langle\bar{\psi}\gamma_5\psi\rangle^{-5}$
   &              &            &            &-0.0828(20) &-0.0700(18)\\
\hline\hline
\end{tabular}
\caption{ The pseudoscalar condensation 
$-\langle\bar{\psi}\gamma_5\psi\rangle^Q$ for a fixed topological charge $Q$ 
at $\beta=1.0$.}
\label{tab:psigammapsi}
\end{table*}

\begin{table*}[p]
\begin{tabular}{cccccc}
\hline
\hline
 & $m=0.1$ & $m=0.15$ & $m=0.2$ & $m=0.25$ & $m=0.3$ \\\hline
 $-\langle\bar{\psi}\psi\rangle^{\theta=0}$    
& 0.1338(93) & 0.150(12) & 0.1678(88) & 0.182(15) & 0.193(10) \\
 $-\langle\bar{\psi}\psi\rangle^{\theta=0.1\pi}$  
& 0.1309(90) & 0.149(12) & 0.1667(87) & 0.181(15) & 0.193(10)\\
 $-\langle\bar{\psi}\psi\rangle^{\theta=0.2\pi}$  
& 0.1219(87) & 0.146(13) & 0.1632(97) & 0.177(19) & 0.190(14)\\
 $-\langle\bar{\psi}\psi\rangle^{\theta=0.3\pi}$  
& 0.107(10)  & 0.140(14) & 0.155(17)  & 0.166(47) & 0.178(56)\\
\hline\hline
\end{tabular}
\caption{The scalar condensation
 $-\langle\bar{\psi}\psi\rangle^{\theta}$ 
in $\theta$ vacuum at $\beta=1.0$.}
\label{tab:psipsitheta}
\end{table*}

\begin{table*}[p]
\begin{tabular}{cccccc}
\hline
\hline
 & $m=0.1$ & $m=0.15$ & $m=0.2$ & $m=0.25$ & $m=0.3$ \\\hline
 $i\langle\bar{\psi}\gamma_5\psi\rangle^{\theta=0}$    
& 0.0(0)     & 0.0(0)     & 0.0(0)     & 0.0(0)     & 0.0(0)\\
 $i\langle\bar{\psi}\gamma_5\psi\rangle^{\theta=0.1\pi}$  
& 0.0122(10) & 0.0147(28) & 0.0166(12) & 0.0180(16) & 0.0159(11)\\
 $i\langle\bar{\psi}\gamma_5\psi\rangle^{\theta=0.2\pi}$  
& 0.0229(20) & 0.0290(46) & 0.0343(25) & 0.0398(42) & 0.0373(29)\\
 $i\langle\bar{\psi}\gamma_5\psi\rangle^{\theta=0.3\pi}$  
& 0.0306(28) & 0.0422(77) & 0.0521(49) & 0.0739(16) & 0.090(15)\\
\hline\hline
\end{tabular}
\caption{The pseudoscalar condensation 
$i\langle\bar{\psi}\gamma_5\psi\rangle^{\theta}$ in $\theta$ vacuum 
at $\beta=1.0$.}
\label{tab:psigammapsitheta}
\end{table*}

\begin{table*}[p]
\begin{tabular}{lccc}
\hline
\hline
 & $m=0.1$ & $m=0.2$ & $m=0.3$ \\\hline
 $\langle\eta^{\dagger}(\infty)\eta(0)\rangle^0$
   & 0.00513(41) & 0.00271(50) & 0.00162(36) \\
 $\langle\eta^{\dagger}(\infty)\eta(0)\rangle^1$   
   & 0.00010(56) & 0.00043(82) & 0.00119(43)\\
 $\langle\eta^{\dagger}(\infty)\eta(0)\rangle^{-1}$
   & 0.0005(36) &  0.00030(48) & -0.07e-3(63)\\
 $\langle\eta^{\dagger}(\infty)\eta(0)\rangle^2$
   &-0.0151(12) & -0.00397(97) & -0.49e-3(54) \\
 $\langle\eta^{\dagger}(\infty)\eta(0)\rangle^{-2}$
   &-0.0184(11) & -0.00399(71) &  0.48e-3(44)\\
 $\langle\eta^{\dagger}(\infty)\eta(0)\rangle^3$
   &-0.0489(33) & -0.0149(57)  & -0.00224(59)\\
 $\langle\eta^{\dagger}(\infty)\eta(0)\rangle^{-3}$
   &-0.0418(19) & -0.0110(18)  & -0.00658(68)\\
 $\langle\eta^{\dagger}(\infty)\eta(0)\rangle^4$
   &-0.0839(23) & -0.0242(10)  & -0.0093(19)\\
 $\langle\eta^{\dagger}(\infty)\eta(0)\rangle^{-4}$
   &-0.0921(27)  &-0.0271(18)  & -0.0143(20)\\
 $\langle\eta^{\dagger}(\infty)\eta(0)\rangle^5$
   &             &             & -0.01476(84)\\
 $\langle\eta^{\dagger}(\infty)\eta(0)\rangle^{-5}$
   &             &             & -0.02055(79)\\
\hline\hline
\end{tabular}
\caption{
The long-range correlation 
$\langle\eta^{\dagger}(\infty)\eta(0)\rangle^Q$ at $\beta=1.0$ 
and $m=0.1, 0.2 , 0.3$.
in each topological sector.}
\label{tab:longcrr}
\end{table*}

\begin{table*}[p]
\begin{tabular}{lccc}
\hline
\hline
 & $m=0.1$& $m=0.15$& $m=0.2$\\\hline
$\langle\eta^{\dagger}(\infty)\eta(0)\rangle^{\theta=0}$     
& 0.03e-3(74)&-0.92e-3(46) &-0.00181(48)\\
$\langle\eta^{\dagger}(\infty)\eta(0)\rangle^{\theta=0.1\pi}$
& 0.83e-3(72)&-0.91e-3(43) &-0.87e-3(38)\\
$\langle\eta^{\dagger}(\infty)\eta(0)\rangle^{\theta=0.2\pi}$
& 0.00304(71)&0.00238(47)  &0.00232(42)\\
$\langle\eta^{\dagger}(\infty)\eta(0)\rangle^{\theta=0.3\pi}$
& 0.00607(72)&0.00630(75)  &0.0089(16)\\
$\langle\eta^{\dagger}(\infty)\eta(0)\rangle^{\theta=0.4\pi}$
& 0.00889(75)&0.0112 (15)  &0.0180(45)\\
$\langle\eta^{\dagger}(\infty)\eta(0)\rangle^{\theta=0.5\pi}$
& 0.01048(88)&0.0171 (35)  &0.0143(49)\\
\hline\hline
\end{tabular}
\caption{
The long-range correlation 
$\langle\eta^{\dagger}(\infty)\eta(0)\rangle^{\theta}$ at $\beta=1.0$
 and $m=0.1, 0.15, 0.2$.}
\label{tab:longcrrtheta}
\end{table*}


\begin{thebibliography}{9}

\bibitem{Ginsparg:1981bj}
P.~H.~Ginsparg and K.~G.~Wilson,
Phys.\ Rev.\ D {\bf 25}, 2649 (1982).


\bibitem{Fukaya:2003ph}
H.~Fukaya and T.~Onogi,
Phys.\ Rev.\ D {\bf 68}, 074503 (2003).


\bibitem{Schwinger:tp}
J.~S.~Schwinger,
Phys.\ Rev.\  {\bf 128}, 2425 (1962).

\bibitem{Coleman:1975pw}
S.~R.~Coleman, R.~Jackiw and L.~Susskind,
Annals Phys.\  {\bf 93}, 267 (1975).

\bibitem{Coleman:1976uz}
S.~R.~Coleman,
Annals Phys.\  {\bf 101}, 239 (1976).

\bibitem{Smilga:1996pi}
A.~V.~Smilga,
Phys.\ Rev.\ D {\bf 55}, 443 (1997)

\bibitem{Hetrick:1995wq}
J.~E.~Hetrick, Y.~Hosotani and S.~Iso,
Phys.\ Lett.\ B {\bf 350}, 92 (1995).

\bibitem{Rodriguez:1996zj}
R.~Rodriguez and Y.~Hosotani,
Phys.\ Lett.\ B {\bf 375}, 273 (1996).


\bibitem{Hosotani:1998za}
Y.~Hosotani and R.~Rodriguez,
J.\ Phys.\ A {\bf 31}, 9925 (1998).

\bibitem{Durr:2003xs}
S.~D\"urr and C.~Hoelbling,
arXiv:hep-lat/0311002.

\bibitem{Durr:2000gi}
S.~D\"urr,
Phys.\ Rev.\ D {\bf 62}, 054502 (2000).

\bibitem{Elser:2001pe}
S.~Elser,
arXiv:hep-lat/0103035.

\bibitem{Elser:1996tb}
S.~Elser and B.~Bunk,
Nucl.\ Phys.\ Proc.\ Suppl.\  {\bf 53}, 953 (1997).

\bibitem{Kiskis:2000sb}
J.~Kiskis and R.~Narayanan
Phys.\ Rev.\ D {\bf 62}, 054501 (2000).

\bibitem{Chandrasekharan:1998em}
S.~Chandrasekharan,
Phys.\ Rev.\ D {\bf 59}, 094502 (1999).

\bibitem{Gattringer:1995du}
C.~Gattringer,
Phys.\ Rev.\ D {\bf 53}, 5090 (1996).

\bibitem{deForcrand:1997fm}
P.~de Forcrand, J.~E.~Hetrick, T.~Takaishi and A.~J.~van der Sijs,
Nucl.\ Phys.\ Proc.\ Suppl.\  {\bf 63} (1998) 679.

\bibitem{Vranas:1997da}
P.~M.~Vranas,
Phys.\ Rev.\ D {\bf 57}, 1415 (1998).

\bibitem{Giusti:2001cn}
L.~Giusti, C.~Hoelbling and C.~Rebbi,
Phys.\ Rev.\ D {\bf 64}, 054501 (2001).

\bibitem{Wiese:1988qz}
U.~J.~Wiese,
Nucl.\ Phys.\ B {\bf 318}, 153 (1989).

\bibitem{Hassan:1994wy}
A.~S.~Hassan, M.~Imachi and H.~Yoneyama,
Prog.\ Theor.\ Phys.\  {\bf 93}, 161 (1995).

\bibitem{Hassan:1995dn}
A.~S.~Hassan, M.~Imachi, N.~Tsuzuki and H.~Yoneyama,
Prog.\ Theor.\ Phys.\  {\bf 94}, 861 (1995).

\bibitem{Imachi:1997hf}
M.~Imachi, T.~Kakitsuka, N.~Tsuzuki and H.~Yoneyama,
arXiv:hep-lat/9702018.

\bibitem{Plefka:1996tz}
J.~C.~Plefka and S.~Samuel,
Phys.\ Rev.\ D {\bf 56}, 44 (1997).

\bibitem{Azcoiti:2002vk}
V.~Azcoiti, G.~Di Carlo, A.~Galante and V.~Laliena,
Phys.\ Rev.\ Lett.\  {\bf 89}, 141601 (2002).

\bibitem{D'Elia:2003gr}
M.~D'Elia,
Nucl.\ Phys.\ B {\bf 661}, 139 (2003).

\bibitem{Luscher:1998du}
M.~L\"uscher, 
Nucl.\ Phys.\ B {\bf 549},  295 (1999).

\bibitem{Kaplan}
D.~B.~Kaplan,
Phys.\ Lett.\ B {\bf 288}, 342 (1992).

\bibitem{Shamir}
Y. Shamir,
Nucl.\ Phys.\ B {\bf 406}, 90 (1993).\\ 
V.~Furman and Y.~Shamir,
Nucl.\ Phys.\ B {\bf 439}, 54 (1995).



\bibitem{Venkataraman:1997xi}
L.~Venkataraman and G.~Kilcup,
arXiv:hep-lat/9711006.

\bibitem{McNeile:2000hf}
C.~McNeile and C.~Michael  [UKQCD Collaboration],
Phys.\ Lett.\ B {\bf 491}, 123 (2000)
[Erratum-ibid.\ B {\bf 551}, 391 (2003)].
%

\bibitem{Struckmann:2000bt}
T.~Struckmann {\it et al.}  [TXL Collaboration],
Phys.\ Rev.\ D {\bf 63}, 074503 (2001).

\bibitem{Lesk:2002gd}
V.~I.~Lesk {\it et al.}  [CP-PACS Collaboration],
Phys.\ Rev.\ D {\bf 67}, 074503 (2003).



\bibitem{ref:noise}
K.~Bitar {\it et al.}, Nucl. Phys. {\bf B313}, 348 (1989).\\
H.~R.~Fiebig and R.~M.~Woloshyn, Phys. Rev. D {\bf 42}, 3520 (1990).

\bibitem{ref:kuramashi}
Y.~Kuramashi {\it et al.}, Phys. Rev. Lett. {\bf 72}, 3448 (1994).\\
M.~Fukugita {\it et al.}, Phys. Rev. D {\bf 51}, 3952 (1995).




\bibitem{Gausterer:1988tg}
H.~Gausterer, J.~Potvin, S.~Sanielevici and P.~Woit,
Phys.\ Lett.\ B {\bf 233}, 439 (1989).

\bibitem{Bitar:1991wr}
K.~M.~Bitar {\it et al.},
Phys.\ Rev.\ D {\bf 44}, 2090 (1991).
%

\bibitem{Kuramashi:1993mv}
Y.~Kuramashi, M.~Fukugita, H.~Mino, M.~Okawa and A.~Ukawa,
Phys.\ Lett.\ B {\bf 313}, 425 (1993).


\bibitem{Hart:2001pj}
A.~Hart and M.~Teper  [UKQCD Collaboration],
Phys.\ Lett.\ B {\bf 523}, 280 (2001).
%

\bibitem{Bali:2001gk}
G.~S.~Bali {\it et al.}  [TXL Collaboration],
Phys.\ Rev.\ D {\bf 64}, 054502 (2001).


\bibitem{Hasenfratz:2001wd}
A.~Hasenfratz,
Phys.\ Rev.\ D {\bf 64}, 074503 (2001).

\bibitem{Alles:2000cg}
B.~Alles, M.~D'Elia and A.~Di Giacomo,
Phys.\ Lett.\ B {\bf 483}, 139 (2000).

\bibitem{Bernard:2003gq}
C.~Bernard {\it et al.},
Phys.\ Rev.\ D {\bf 68}, 114501 (2003).

\bibitem{:2004ij}
UKQCD and A.~Hart  [QCDSF Collaborations],
arXiv:hep-lat/0401015.



\end{thebibliography}
\end{document}